\documentclass[%
reprint,
 superscriptaddress,
 amsmath,
 amssymb,
aps,
floatfix,
]{revtex4-2}

\usepackage{graphicx}
\usepackage{dcolumn}
\usepackage{bm}
\usepackage{hyperref}
\usepackage{gensymb}
\usepackage{xcolor}
\usepackage{algpseudocode}
\usepackage{algorithm}
\usepackage{lipsum}  
\usepackage{textgreek}

\hypersetup{
    colorlinks=true,
    linkcolor=blue,
    citecolor=blue, 
    urlcolor=black
    }

\begin{document}

\title{Geometric scaling of laser-driven proton focusing from hemispherical foils}

\author{J. Griff-McMahon}
\email{jgriffmc@pppl.gov}
\affiliation{Department of Astrophysical Sciences, Princeton University, Princeton, New Jersey 08544, USA}
\affiliation{Princeton Plasma Physics Laboratory, Princeton, New Jersey 08540, USA}

\author{X. Vaisseau}
\affiliation{Focused Energy Inc., Redwood City, California 94063, USA}

\author{W. Fox}
\affiliation{Department of Astrophysical Sciences, Princeton University, Princeton, New Jersey 08544, USA}
\affiliation{Princeton Plasma Physics Laboratory, Princeton, New Jersey 08540, USA}
\affiliation{Department of Physics, University of Maryland, College Park, Maryland 20742, USA}

\author{K. Lezhnin}
\affiliation{Princeton Plasma Physics Laboratory, Princeton, New Jersey 08540, USA}

\author{K. Bhutwala}
\affiliation{Princeton Plasma Physics Laboratory, Princeton, New Jersey 08540, USA}

\author{R. Nedbailo}
\affiliation{University of Texas, Austin, Texas 78701, USA}

\author{V. Ospina-Bohórquez}
\affiliation{Focused Energy Inc., Redwood City, California 94063, USA}

\author{T. Karpowski}
\affiliation{Focused Energy GmbH, 64293 Darmstadt, Germany}

\author{P. K. Patel}
\affiliation{Focused Energy Inc., Redwood City, California 94063, USA}

\author{S. Malko}
\affiliation{Princeton Plasma Physics Laboratory, Princeton, New Jersey 08540, USA}

\begin{abstract}
We systematically characterize the focusing behavior of laser-driven proton beams from hemispherical targets of various diameters using mesh radiography. The proton focal location is inferred to be near the geometrical center for the smallest tested hemisphere ($\Psi=D_{hemi}/D_{Laser}=6.1$). However, larger hemispheres ($\Psi=14.6$) degrade the focusing behavior and behave more like flat foils with focal location significantly inside the hemisphere. We also infer a tight virtual focus of $9\pm3~\mu$m through a mesh transition analysis.
\end{abstract}

\maketitle

\section{Introduction}
Tightly focused beams of multi-MeV proton are a powerful tool for creating and probing matter in extreme states. They can isochorically heat solid-density materials to the warm-dense matter regime \cite{patel_isochoric_2003, bailly-grandvaux_creation_2025} and are central to the advanced inertial confinement fusion scheme of proton fast ignition (PFI) \cite{roth_fast_2001}. Unlike central hot spot ignition, PFI separates the compression and heating stages in inertial confinement fusion, offering potential improvements over central hotspot ignition through higher target gains and relaxed symmetry requirements. Long-pulse lasers precompress the fuel to a relatively cold and dense state, whereby a high-fluence proton beam isochorically heats and ignites the assembly. From several design studies, ignition requires a proton beam with $\sim20~$kJ of energy in a temperature from 4 to 8 MeV, delivered within a beam radius of $\sim20~\mu$m and in less than 20 ps before the fuel disassembles \cite{roth_fast_2001,atzeni_first_2002,tabak_fast_2006,honrubia_ion_2015}. 

To achieve these requirements, high-intensity lasers irradiate a curved, hemispherical foil or ``hemi" to generate a focused proton beam. This scheme uses the well known Target Normal Sheath Acceleration (TNSA) mechanism \cite{wilks_energetic_2001} in which relativistic electrons are energized from the laser interaction and create a strong electrostatic sheath field on the rear side of the target, capable of accelerating protons to MeV energies. Since protons are predominantly accelerated in the target-normal direction on the rear of the target, using a curved target geometrically focuses protons towards a common focal spot. In the final PFI design, the hemi is placed inside a hollow cone, primarily to protect the hemi, but ultimately plays a role in proton focusing \cite{bartal_focusing_2012,honrubia_ion_2015,qiao_dynamics_2013,mcguffey_focussing_2020}. In the context of PFI, the beam focal spot size and the focal location are critical. 

Design studies \cite{honrubia_ion_2015} identify an optimal beam radius of 15 to 20$~\mu$m at 300 g$/$cm$^3$ fuel density, which balances two competing effects. Smaller beams produce a stronger hydrodynamic response and suffer from range lengthening by rapidly heating and rarefying the deposition channel so that protons arriving at later times deposit their energy deeper into the fuel assembly. However, larger beams distribute the energy over a greater volume, lowering the achieved temperature. Likewise, the axial location of the beam focus should ideally coincide with the densest part of the fuel assembly to minimize the ignition energy required in the proton beam. The axial location also balances two competing requirements: the hemi target must be placed sufficiently far from the fusion capsule to prevent the imploding target from disrupting the protective cone structure used in fast-ignition designs, yet close enough to limit temporal dispersion from a poly-energetic beam.

\begin{figure} [b]
	\includegraphics[width=\linewidth]{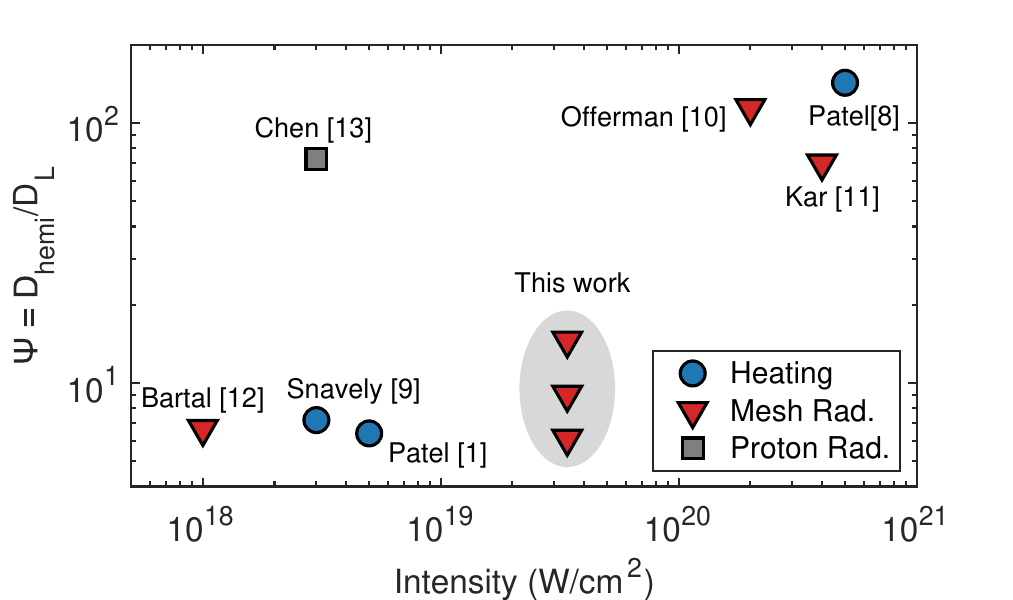}
	\caption{Experimental parameter space of dimensionless focusing geometry $\Psi=D_{hemi}/D_L$ and laser intensity from Refs. \cite{patel_isochoric_2003,patel_integrated_2005,snavely_laser_2007,offermann_characterization_2011,kar_ballistic_2011,bartal_focusing_2012,chen_focusing_2012} and this work. Refs. \cite{kar_ballistic_2011,chen_focusing_2012} used semi-cylinders. The markers illustrate the different experimental characterization techniques of heating a secondary sample, mesh radiography, and side-on proton radiography.}
	\label{fig:FocusingParameterSpace}
\end{figure}

Many prior experiments have shown that curved targets lead to improved proton focusing and heating of a secondary sample, as compared to flat targets \cite{patel_isochoric_2003, patel_integrated_2005,snavely_laser_2007, offermann_characterization_2011, kar_ballistic_2011, bartal_focusing_2012, chen_focusing_2012}. Two important parameters in these experiments are the laser intensity which roughly controls the electron temperature (and therefore the TNSA electric fields since $E_{TNSA}\propto T_e \propto \sqrt{I_L}$) and the dimensionless focusing geometry defined as the ratio of the hemi diameter to the laser spot diameter ($\Psi=D_{hemi}/D_{Laser}$) \cite{ospina-bohorquez_scaling_2025}. Figure \ref{fig:FocusingParameterSpace} shows prior focusing experiments plotted in this parameter space.

Patel \emph{et al.} \cite{patel_isochoric_2003} first demonstrated heating of a secondary sample from a hemi, resulting in a heating area 16$\times$ smaller than from a flat target. Later, Snavely \emph{et al.} \cite{snavely_laser_2007} performed an experiment at higher energy and showed that optimal heating of a secondary sample occurs when the sample is placed roughly 1 radius downstream of the hemi geometrical center. Several other experiments characterized the focusing dynamics using mesh radiography and inferred that proton trajectories were curved and bent outwards around the focus from radial electric fields \cite{offermann_characterization_2011,bartal_focusing_2012}. 
The experiments established proton focusing as a powerful tool, however there are several open questions surrounding its robustness, scalability, and the key parameters for optimization.
For one, these experiments inferred their conclusions from a small number of shots (often $<5$), which are susceptible to shot-to-shot variations of laser parameters and pointing stability between the hemi center and laser focal spot. Obtaining statistical measurements in these systems is vital to understanding the hemi focusing dynamics. Secondly, many of these works \cite{chen_focusing_2012,offermann_characterization_2011,kar_ballistic_2011,patel_integrated_2005} performed experiments with large values of $\Psi>70$ that may have different focusing behavior than small $\Psi$. 
Recent simulation studies have identified small $\Psi$ ($\lesssim15$) as a key for optimized focusing \cite{kemp_laser--proton_2024} and showed that best focusing occurs at $\Psi = 6$ to 8.5 for a 40 fs, 18 J, $4\times10^{19}~$W/cm$^2$ drive \cite{ospina-bohorquez_scaling_2025}. Furthermore, while PFI requires proton beams from multi-kJ ultrashort lasers, such drivers do not yet exist, preventing direct experimental validation of ignition-scale focusing. Instead, medium-scale, high-repetition-rate facilities offer a path forward: they enable collection of the large datasets needed to characterize variability and focusing dynamics in the lower-energy regime, and provide crucial benchmarks for particle-in-cell (PIC) codes. These codes, once validated, can then be used to reliably extrapolate to ignition-relevant conditions. Beyond fast ignition, systematic optimization and characterization of proton focusing are equally critical for generating well-controlled warm dense matter (WDM) samples, enabling improved understanding of fundamental material properties like proton transport \cite{malko_proton_2022} and equation of state \cite{hoarty_equation_2012}.

In this work, we performed a parametric study of hemisphere diameter on proton focusing characteristics at the ALEPH laser at Colorado State University. Enhanced shot rate targetry and small hemis were fielded to diagnose the proton beam dynamics over 70 shots. $\Psi=D_{hemi}/D_{L}$ was varied between 6 and 15. Mesh radiography characterized the focal location at several mesh distances to inform both the virutal and physical focal locations. The smallest hemisphere $(\Psi=6.1)$ focused protons near the geometric center, while the largest hemisphere $(\Psi=14.6)$ focused deeper inside the target and has degraded focusing. The proton virtual focal spot size was inferred to be $\sim9~\mu$m through a mesh transition analysis. In addition, the proton beam pointing was more erratic for smaller hemis as laser pointing stability had a stronger effect.

Section \ref{sec:setup} introduces the experimental setup. Sections \ref{sec:focal_location} and \ref{sec:focal_size} characterize the focal location and focal spot size using a mesh radiography analysis. Section \ref{sec:beam_pointing} studies the beam pointing and divergence behavior and Sec. \ref{sec:conclusion} provides a discussion and comparison to previous experiments.

\section{Experimental Setup} \label{sec:setup}

\begin{figure}
	\includegraphics[width=\linewidth]{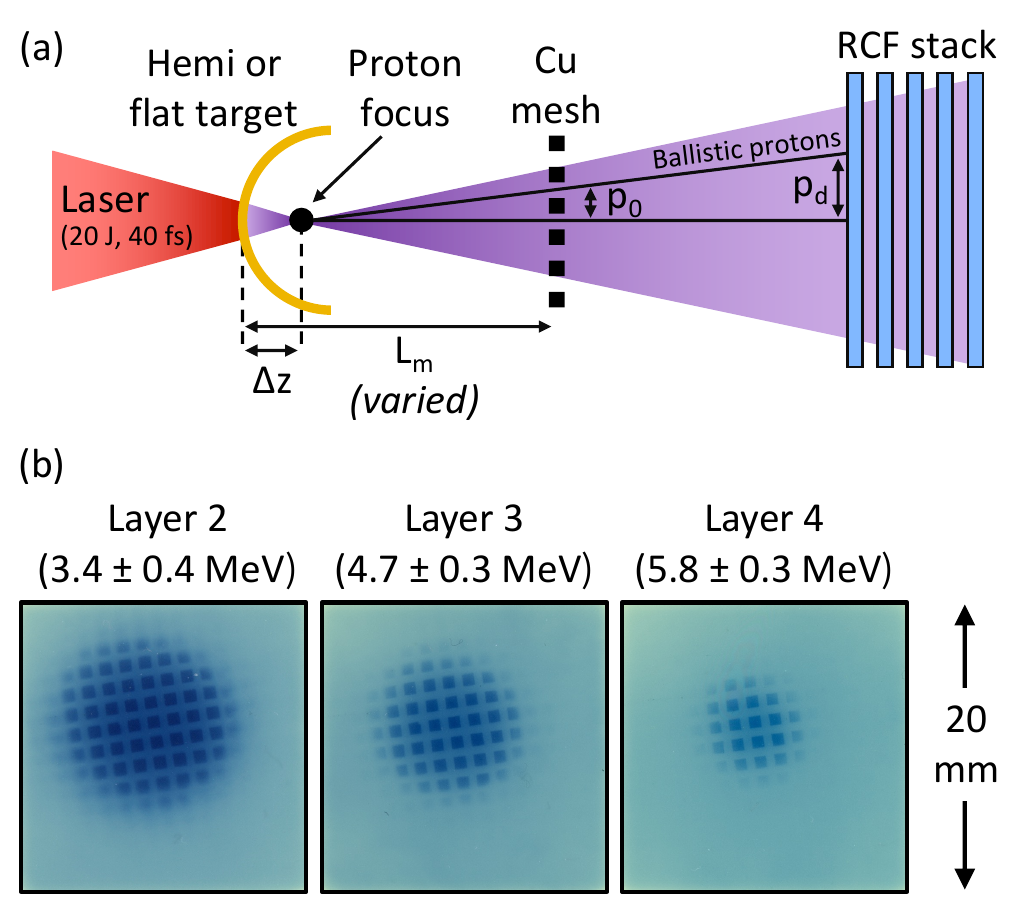}
	\caption{(a) Mesh radiography setup used to characterize the proton focusing. $p_0$ is the mesh pitch and $p_d$ is the magnified pitch on the detector. (b) RCF images from a single shot showing different layers and energy resolution.}
	\label{fig:setup_data}
\end{figure}
The experiment was performed at the ALEPH laser at Colorado State University and the setup is shown in Fig. \ref{fig:setup_data}(a). A short-pulse laser irradiates a 10 $\mu$m-thick Au target, which is either a hemi or flat foil, to produce a beam of TNSA protons on the rear side of the target. The hemi diameter was varied between 220, 325, and $525~\mu$m. An enhanced shot rate target assembly was used to house the targets, enabling pre-characterization  and automatic alignment of hemis \cite{bauer_development_2025}. The laser was characterized by $19.9\pm0.9$ J of energy, 40 fs FWHM pulse duration, 800 nm wavelength, $36\pm4$ $\mu$m FWHM spot size, and peak intensity of $(3.4\pm 0.8)\times10^{19}$ W/cm$^2$ on target. A $25~\mu$m-thick Cu mesh is placed behind the target to imprint a shadow into the proton beam. The distance between the target and the mesh $L_m$ is varied for each target configuration between 280 and $960~\mu$m to gain insight into the proton beam dynamics. Meshes placed closer to the target had a pitch of 400 lines per inch while meshes further away had a finer pitch of 600 lines per inch. The $L_m$ distances are given in Table \ref{tab:MeshDistances} and characterized to within $30~\mu$m. The proton beam fluence is measured with an RCF stack positioned 20 mm away from the target. The stack consisted of 12.5 $\mu$m-thick Al foil and 5 layers of HD-V2 Radiochromic film and provided energy resolution into five proton energy bins with central energies of 1.3, 3.4, 4.7, 5.8, and 6.7 MeV and bandwidths of 0.3 to 0.6 MeV. Figure \ref{fig:setup_data}(b) shows typical RCF images from three different layers in a single shot. A clear mesh imprint is visible and enables high quality data extraction. Over 70 radiography shots were performed during five days of data collection due to the enhanced shot rate. RCF stacks were mounted in a 4x5 frame allowing acquisition of 20 shots without breaking vacuum.

\begin{table}
    \begin{ruledtabular}
    \begin{tabular}{cccc}
    Target Type & $L_{m,1}$ ($\mu$m) & $L_{m,2}$ ($\mu$m) & $L_{m,3}$ ($\mu$m) \\
    \hline
    Flat & 340 & --- & 960\\
    Hemi 525 $\mu$m & --- & 710 & 960\\
    Hemi 325 $\mu$m & 340 & 650 & 960\\
    Hemi 220 $\mu$m & 280 & 600 & 960\\
    \end{tabular}
    \end{ruledtabular}
    \caption{Mesh distances from the hemi apex $L_m$ for each target type.}
    \label{tab:MeshDistances}
\end{table}

\section{Focal location} \label{sec:focal_location}
The mesh is used to infer the longitudinal position of the proton focus based on the observed magnification of the mesh in the RCF image. We assume straight-line trajectories of the proton beam from a source location through the mesh holes onto the detector. Due to this assumption, mesh radiography implicitly measures the location of the virtual proton focus inferred from the secant line between the mesh image on the detector and the physical mesh \cite{borghesi_multi-mev_2004}. This may differ from the actual focus if the proton trajectories are curved. The virtual focus location $\Delta z$ is defined relative to the hemi apex in Fig. \ref{fig:setup_data}(a) and given by
\begin{equation} \label{eq:dz_proton_focus}
    \Delta z = L_m - L_d \frac{p_0}{p_d-p_0}.
\end{equation}
Here, $L_m$ is the distance from the hemi apex to the mesh, $L_d=20~$mm$-L_m$ is the distance from the mesh to the detector, $p_0$ is the mesh pitch, and $p_d$ is the magnified pitch measured on the detector. $p_d$ is found by hand for each shot and layer by locating the mesh crossbars in the RCF images and calculating the average spacing. The mesh location $L_m$ is varied to 2 or 3 positions for each target, given in Table \ref{tab:MeshDistances}.

\begin{figure}
	\includegraphics[width=\linewidth]{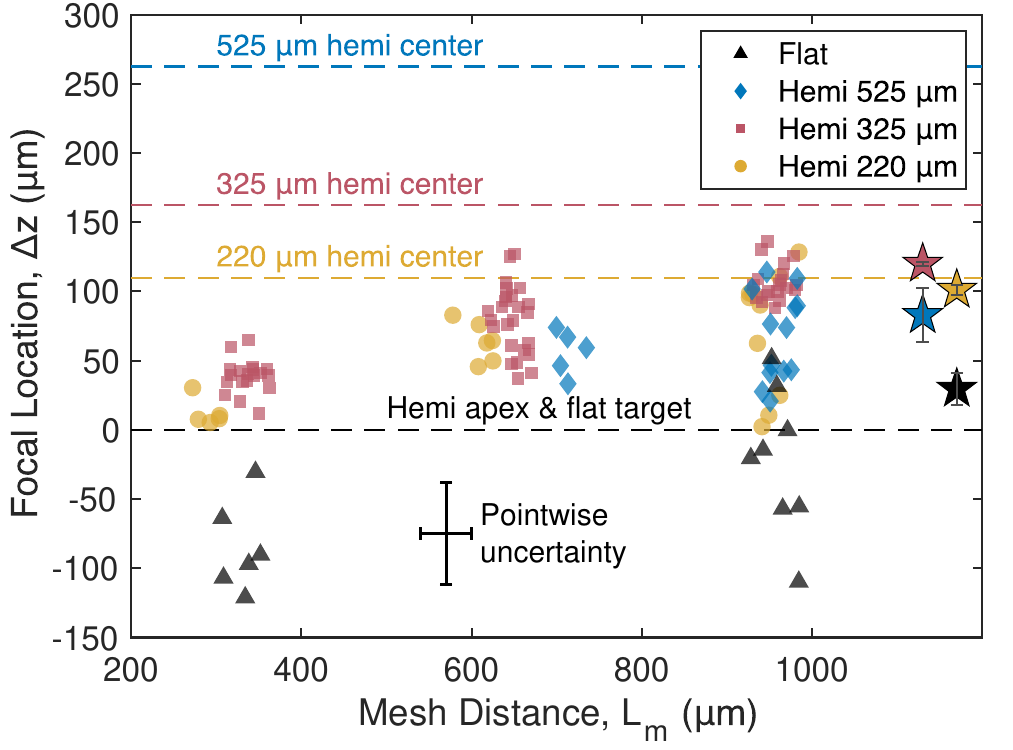}
	\caption{Virtual focal location $\Delta z$ as the mesh distance $L_m$ is varied. For reference, $\Delta z=0$ corresponds to the front surface of the target where the laser hits (hemi apex or flat foil) as shown in Fig. \ref{fig:setup_data}(a). The flat target is shown in black triangles, while the hemis of diameter 525, 325, and 220 $\mu$m are shown in blue diamond, red square, and yellow circle, respectively. RCF layers 2, 3, and 4 are shown. The dashed lines correspond to the geometric center of each hemi target. The stars on the right side are the inferred physical focus from fitting hyperbolic trajectories to the data with errorbars indicating the fit uncertainty. A small amount of horizontal spreading is applied to the datapoints for visibility.}
	\label{fig:dz_vs_distance}
\end{figure}

Figure \ref{fig:dz_vs_distance} shows the virtual focus location for different target types as the mesh distance is varied. Each datapoint corresponds to the inferred virtual focus location for a single layer in one shot. The black data are from flat targets while the blue, red, and yellow  data refer to hemis with diameters of 525, 325, and 220$~\mu$m, respectively. The black dashed line denotes $\Delta z =0$, which is the apex of the hemisphere targets where the laser hits and the location of the flat targets. The different colored dashed lines refer to the geometric center (i.e. radius) of the different sized hemis.

For nearly all shots, we observe that the virtual focal location is less than the hemi radius, corresponding to the region inside the hemi. Surprisingly, we also find that the virtual focal location from the hemi apex is nearly the same for all tested hemi sizes. The flat target produces a virtual focus that is behind the target due to laminar expansion, as demonstrated in previous work \cite{borghesi_multi-mev_2004}.

In addition, there is a clear trend in virtual focus location moving downstream (away from the target) as the mesh is moved further away. This trend was also observed in Offerman \emph{et al.} \cite{offermann_characterization_2011} and can be explained by an evolving proton trajectory that passes through the mesh plane before the beam becomes ballistic. Prior work has shown that protons focused from hemis may not travel in straight lines through the geometric center but are rather curved away from the axis due to the hot electron pressure \cite{offermann_characterization_2011, bartal_focusing_2012, foord_proton_2012, bellei_electron_2012}. The electron pressure sets up a radial electric field $E_r \approx -\nabla P_e/(e n_e)$ that first focuses protons inwards and then switches sign to defocus and curve protons outwards. There are several models for proton focusing including that explored in Bellei \emph{et al.} \cite{bellei_electron_2012} and the isothermal model presented in Offerman \emph{et al.} \cite{offermann_characterization_2011}. However, in this work we assume a hyperbolic trajectory for simplicity which showed good agreement to the isothermal model in \cite{offermann_characterization_2011} and replicates the main characteristics of curved trajectories near the focus and asymptotic evolution away from the focus.

\begin{figure}
    \includegraphics[width=\linewidth]{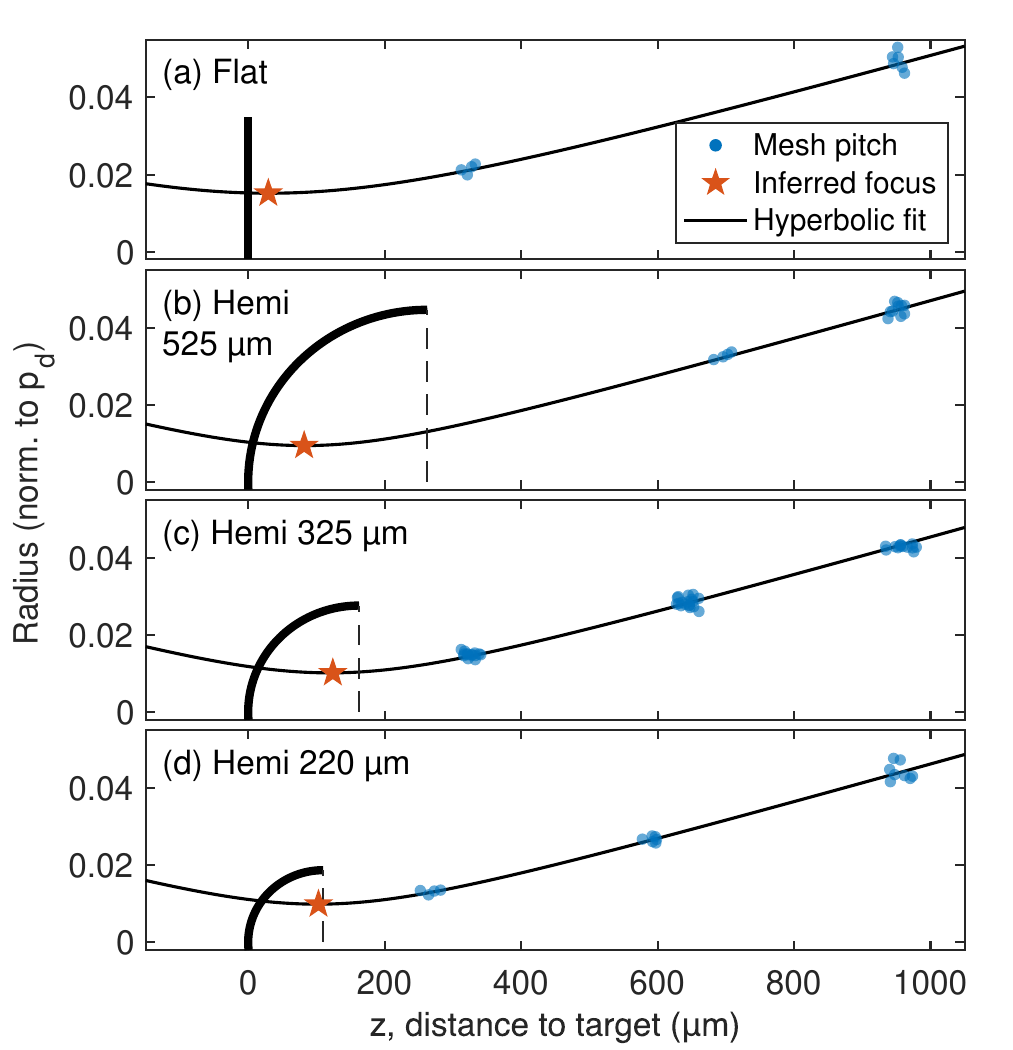}
    \caption{Sample proton trajectories for each target type. The blue dots show the mesh pitch normalized to the measured detector pitch $(p_0/p_d)$ at different mesh positions $L_m$. The data are plotted this way to compare self-similar proton trajectories. A hyperbolic fit is given for each target type (solid line). The red star is the physical focal position based on the minimum of the hyperbolic trajectory. A small amount of horizontal spreading is applied to the data for visibility. Note that the hemis are depicted with a circular aspect ratio but should actually be stretched in the radial direction from the normalization.}
    \label{fig:beam_evol}
\end{figure}

Figure \ref{fig:beam_evol} shows evolving hyperbolic proton trajectory through the mesh data for each target. The radial coordinate is normalized to the measured detector pitch $p_d$ so that data from different mesh locations can be compared on the same footing on self-similar trajectories. This leads to a spread in mesh pitch datapoints $p_0/p_d$ and a datapoint at the RCF plane at $(z=20~\text{mm},~p_d/p_d=1)$. The fitted trajectory assumes the following hyperbolic form
\begin{equation} \label{eqn:proton_hyperbola}
    R(z) = m\sqrt{(z-\Delta z_{phys})^2 + r_0^2}
\end{equation}
with asymptotic slope $m$, physical focal location $\Delta z_{phys}$, and minimum proton beam radius $m r_0$. If $r_0$ and $\Delta z$ are kept constant, then all trajectories have the same asymptotic virtual focus location, regardless of the value of $m$. Experimentally, this manifests as having the same detector pitch everywhere on the image so that different radii trace back to the same focal location within the small angle (paraxial) approximation.

\begin{figure}
    \includegraphics[width=\linewidth]{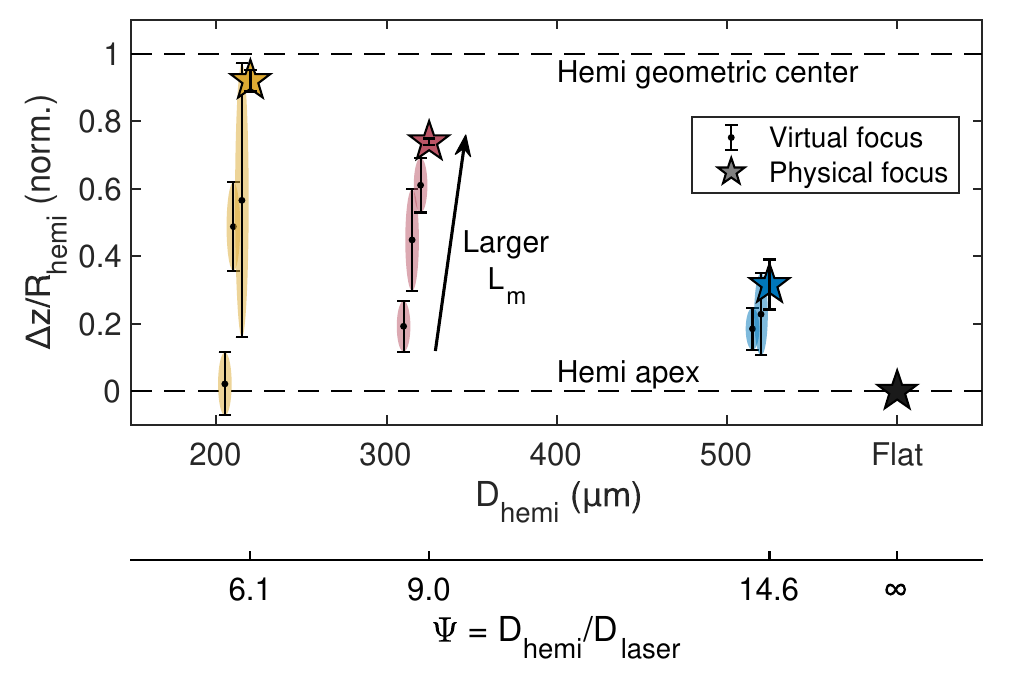}
    \caption{Normalized focal location $\Delta z/R_{hemi}$ for different hemi diameters. Virtual focal locations are shown by shaded ellipses and errorbars containing one standard deviation. The virtual focii have slight horizontal offsets corresponding to different mesh distances $L_m$ and culminating in a star referring to the inferred physical focal, or equivalently the virtual focus as $L_m \to \infty$. The $\Psi=D_{hemi}/D_{Laser}$ value for each target is given in the secondary x-axis.}
    \label{fig:Dz_norm}
\end{figure}

The minimum of each hyperbolic fit is used to infer the proton focal location and is plotted as a star on the right side of Fig. \ref{fig:dz_vs_distance}. These locations are equivalent to the virtual focus if the mesh were placed infinitely far away. All three hemi diameters have similar asymptotic focal locations of $\Delta z_{phys} \approx100~\mu$m from the hemi apex, detailed in Table \ref{tab:FocusLitComp} in Sec. \ref{sec:conclusion}. The flat foil has $\Delta z_{phys} =30 \pm 12~\mu$m, close to the foil surface as expected. The 220 and 325$~\mu$m hemis have small uncertainties in the focal location of less than $4~\mu$m while the 525$~\mu$m hemi has $20~\mu$m uncertainty because there are only two mesh positions for this target type.
Through this hyperbolic fit, we may connect the virtual and physical sources and observe why the virtual source changes with mesh distance. The ratio of the first to second terms in Eq. \eqref{eqn:proton_hyperbola} controls when the trajectory becomes ballistic. Protons are ballistic when they reach the mesh if $L_m-\Delta z _{phys} \gg r_0$. In this case, the virtual focus location will coincide with the physical location. In contrast, if $L_m - \Delta z _{phys} \lesssim r_0$, then the beam is still evolving when it passes through the mesh and the virtual focus will be upstream of the physical source. The hyperbolic fits found $r_0$ to be between 200 and 300$~\mu$m. However, if the proton trajectory is non-hyperbolic, then it is less clear how the physical and virtual focuses are connected. For example, extended channeling such as that observed in Ref. \cite{ospina-bohorquez_scaling_2025} would lead to virtual focusing near the end of the channel.

Figure \ref{fig:Dz_norm} plots the focal location normalized to the hemi radius $\Delta z/R_{hemi}$ for different hemi sizes. A value of $\Delta z/R_{hemi}=1$ corresponds to a focus at the hemi geometric center while a value of $\Delta z/R_{hemi}=0$ is a focus at the hemi apex. The virtual focal locations at intermediate mesh distances $L_m$ are plotted as shaded ellipses and the physical focal location is denoted by a star. Interestingly, a clear trend emerges where the smallest hemi focuses near the geometric center with $\Delta z _{phys} /R_h=0.92$ and the larger hemis focus to points deeper inside the hemisphere with $\Delta z _{phys} /R_h=0.74$ and 0.32 for the 325 and 525$~\mu$m hemis. The focusing from larger hemis (i.e. larger $\Psi$) is degraded and behaves more like a flat foil, as observed in Ref. \cite{ospina-bohorquez_scaling_2025}.

\begin{figure}
	\includegraphics[width=\linewidth]{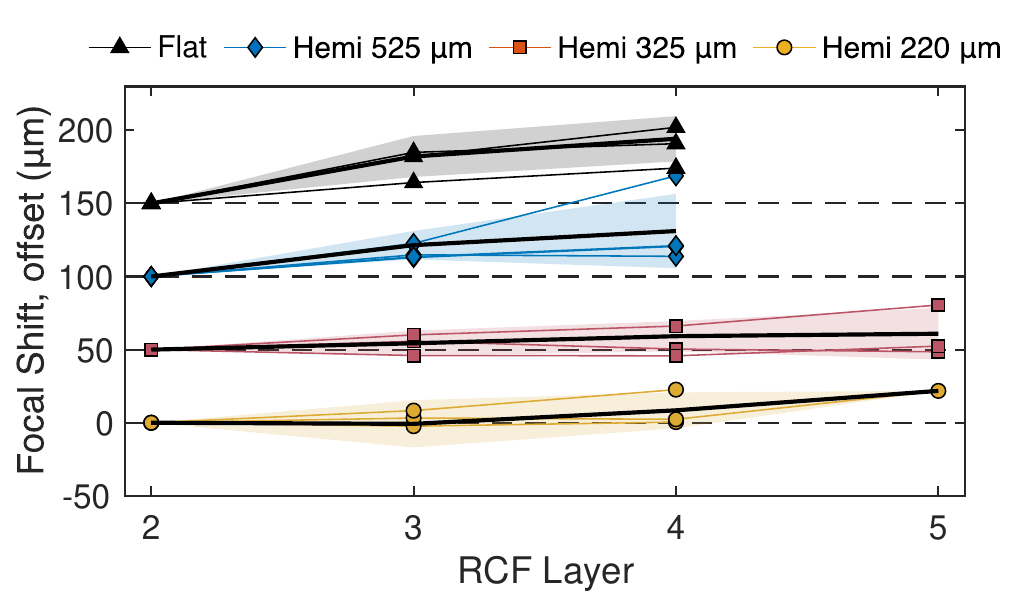}
	\caption{Focal location shift relative to layer 2 for different RCF layers. Different colors correspond to different target types and the plot is offset for each target. The thick black line is the mean value and the shaded region shows $\pm1$ standard deviation. Selected shots are shown as markers where thin lines connect data from a single shot.}
	\label{fig:dz_vs_layer}
\end{figure}

In order to see trends with proton energy, Fig. \ref{fig:dz_vs_layer} shows the virtual focal location shift relative to Layer 2 for different RCF layers. This quantity removes the $\pm30~\mu$m variation in $\Delta z$ that arises from uncertainties in the mesh distance characterization and isolates trends from a single shot. The virtual focus tends to move downstream for greater RCF layers and is more pronounced for hemis with larger diameter and for the flat foil.

\section{Focal spot size} \label{sec:focal_size}
\begin{figure}
	\includegraphics[width=\linewidth]{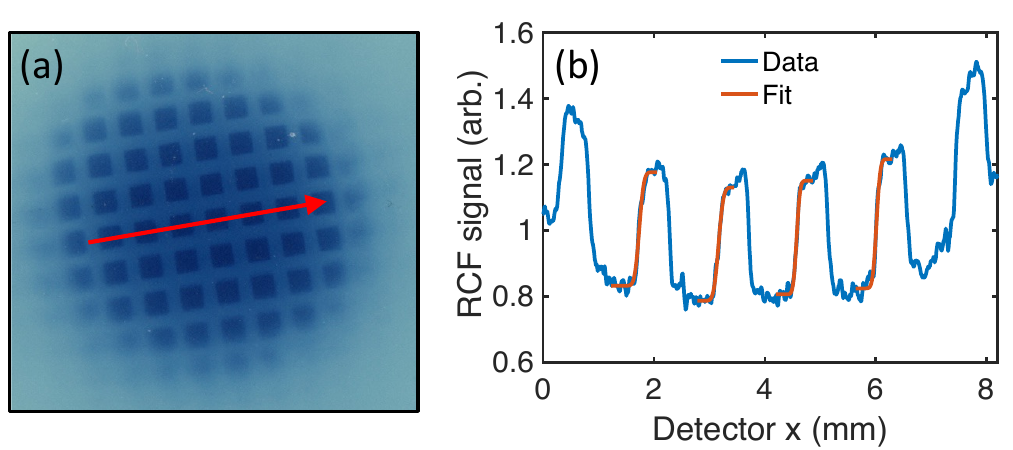}
	\caption{(a) Mesh radiograph with lineout location. (b) Lineout with several error function fits around the mesh transition to infer virtual spot size.}
	\label{fig:focal_spot_overview}
\end{figure}

The transverse proton focal spot size is relevant for proton fast ignition to efficiently heat the target and ensure adequate proton fluence to ignite the compressed fuel \cite{honrubia_ion_2015,atzeni_first_2002}. We measure the virtual focus size using a mesh transition analysis shown in Fig. \ref{fig:focal_spot_overview} and first discussed in Borghesi \emph{et al.} \cite{borghesi_multi-mev_2004}. The sharpness of the mesh transition is directly related to the size of the focus; a larger focus leads to a more blurred image. To extract the focus size, we model the transverse profile $w_s(x)$ as a Gaussian with width $a$. With this assumption, the signal on the detector $I_d(x)$ transitions across a mesh cell according to an error function with spatial scale $Ma$ where $M$ is the mesh magnification in the detector plane.

\begin{align}
    w_s(x) \propto \exp\left(-\frac{x^2}{a^2}\right)\\
    I_d(x) \propto 1+\mathrm{erf}\left(\frac{x}{Ma}\right) 
\end{align}

\begin{figure}
	\includegraphics[width=\linewidth]{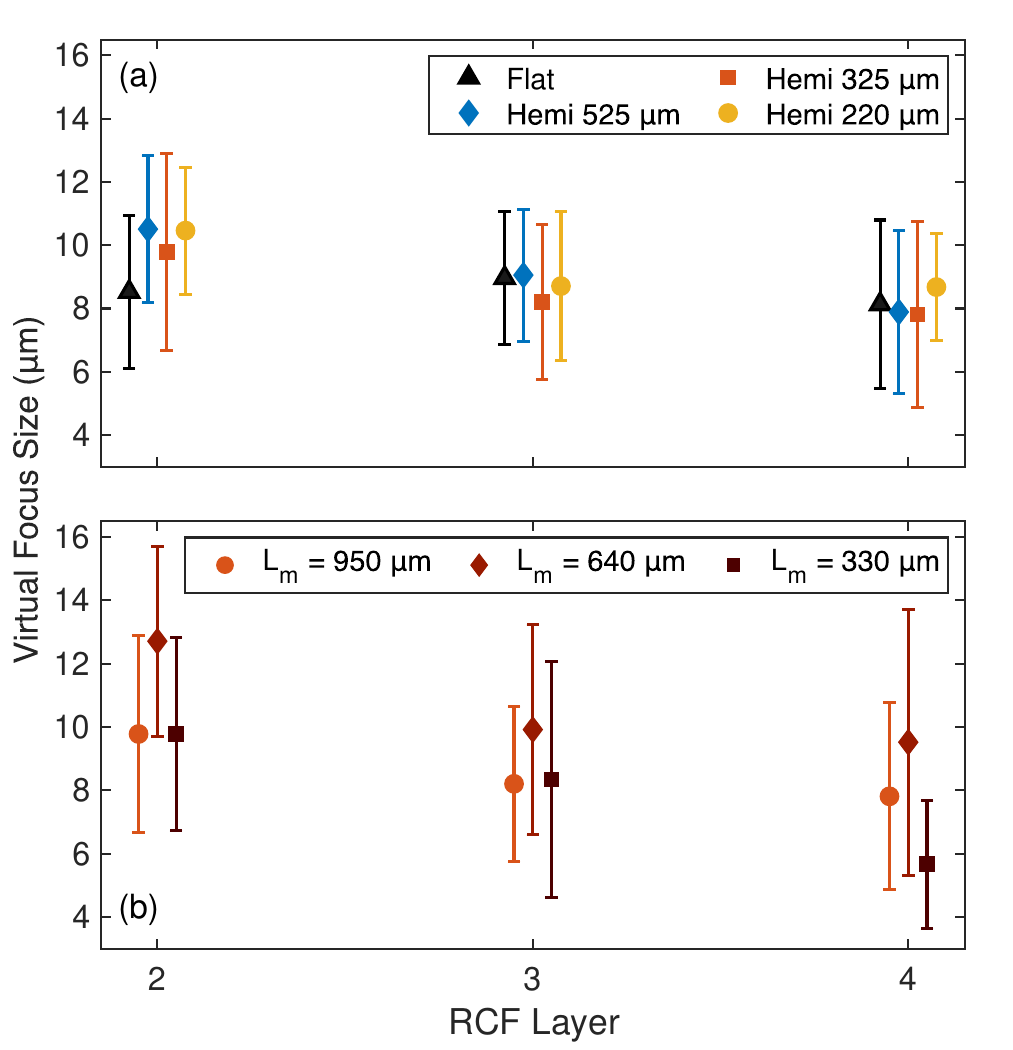}
	\caption{(a) Virtual focus size FWHM vs. layer (energy) for each target type at a fixed mesh distance of 960 $\mu$m. (b) Virtual focus size vs. layer for the 325 $\mu$m hemi, showing different mesh distances, $L_m$.}
	\label{fig:focal_spot_plot}
\end{figure}

25 different shots were analyzed across layers 2 to 4, with 18 transitions fit per image, totaling over 1000 different transitions to characterize the proton focal spot size. Across all the different target types and layers, the mean virtual focal spot FWHM is $9.3 \pm 3 ~\mu$m.  Figure \ref{fig:focal_spot_plot}(a) shows the virtual focal size as a function of the RCF layer for each target type, at constant mesh distance $L_m=960~\mu$m. The proton beam focal spot size decreases slightly for higher RCF layer as expected; the mean size for RCF layer 2 is $9.9\pm 2.6~\mu$m, compared to $8.0\pm 2.6~\mu$m for layer 4. Higher energies are less divergent and therefore exhibit a tighter focus. There is no clear trend between the target types. Figure \ref{fig:focal_spot_plot}(b) shows measurements of the focal size for the $325~\mu$m diameter hemi at different mesh distances, $L_m$. There is no clear trend in the focal size with mesh distance, suggesting that mesh distance does not impact the virtual focal size to the same degree as it does the virtual focal location. With this being said, the virtual focus size is a lower bound to the physical focus size and the physical focus may be larger.

One important note is that the magnification is given by the ratio between the source-to-detector distance and the source-to-mesh distance, $M=(L_d+L_m-\Delta z)/(L_m-\Delta z)$, which includes the focal location $\Delta z$ in the calculation. For simplicity in this analysis, we have assumed $\Delta z=0$ so that the source location is at the front surface of the target at the hemi apex, which would produce an estimated magnification $M_{est}= (L_d+L_m)/L_m$. Since hemis focus protons downstream of the apex, $M_{est}$ underestimates the true magnification and therefore overestimates the virtual focus size for a measured transition size $M_{est}a$. Inserting the inferred focal location of $\Delta z \approx~100~\mu$m decreases the spot size by 11\%, 16\%, and 30\% for mesh distances of 950, 640, and 330$~\mu$m, respectively. However, the fact that there is no trend in virtual focal size with mesh distance [Fig. \ref{fig:focal_spot_plot}(b)] suggests that our estimate for $\Delta z = 0$ is adequate.

\section{Beam pointing and divergence} \label{sec:beam_pointing}

\begin{figure}
	\includegraphics[width=\linewidth]{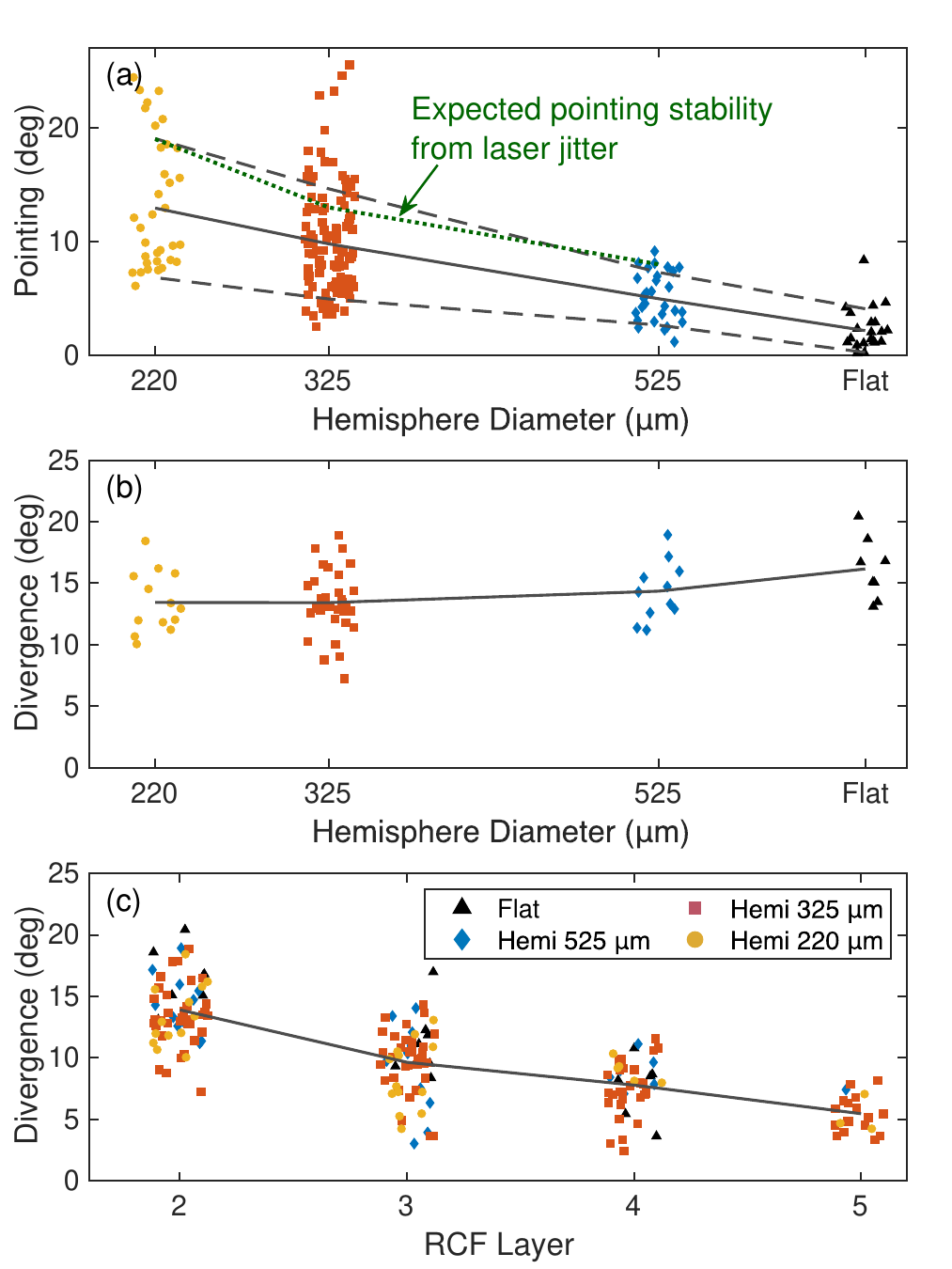}
	\caption{(a) Beam pointing relative to the center of the RCF image for different hemi diameters with the mean value in solid line and $\pm$ standard deviation in dashed lines. The dotted green line shows the expected proton beam pointing from laser jitter, assuming a laser mispoint of one spot size and target normal proton beam propagation. (b) Beam divergence HWHM for RCF layer 2. (c) Beam divergence HWHM for different RCF layers. A small amount of horizontal spreading is applied to all of the data for visibility.}
	\label{fig:pointing_divergence}
\end{figure}

The proton beam pointing and divergence are important properties in the context of proton fast ignition. In contrast to TNSA from a flat foil, the structured nature of a  hemispherical target is relatively sensitive to laser mispointings, especially for small $\Psi$ values. From simulations, laser mispointings leads to a focused proton beam that propagates mainly in the direction normal to where the laser hits, resulting in a beam pointing that is mirrored across the hemi axis \cite{chen_focusing_2012,Lezhnin2024ConcaveTargets}.

Here, we quantify the proton beam pointing and divergence across 66 shots and 4 layers. Figure \ref{fig:pointing_divergence}(a) shows the beam pointing relative to the proton beam centroid from the flat target (close to the center of the RCF). The proton beam is centered on the RCF more frequently as the hemi diameter is increased. This is consistent with laser mispointings becoming less important for the larger hemis. If we take a simple estimate of the laser jitter to be $36~\mu$m (equivalent to one spot size) with geometrically focused protons, then a ballistically focused proton beam would be directed at 19$\degree$, 13$\degree$, and 8$\degree$ from the center of the RCF for the 220, 325, and 525 $\mu$m hemis, respectively. These values are roughly consistent with one standard deviation above the mean of the pointing data, shown as dotted lines.

Figure \ref{fig:pointing_divergence}(b) shows the beam divergence HWHM for RCF layer 2 for different hemi diameters. The divergence is roughly constant at $\sim15\degree$ with a small upward trend of a few degrees from the $220~\mu$m target to the flat target. The divergence measures the asymptotic behavior of the proton beam downstream of the proton focus. If the assumption of a hyperbolic trajectory holds near the proton focus, the beam divergence can be combined with the hyperbolic fit to recover an estimate of the physical proton focal spot size using Eq. \eqref{eqn:proton_hyperbola}. For a divergence of 15\degree, the minimum proton beam radius $mr_0$ is estimated to be 50 to 80$~\mu$m for all targets. This is 3-4$\times$ larger than the laser spot which has a FWHM of 36$~\mu$m. Although it is unclear if the hyperbolic model is valid close to the focus, this result suggests that the virtual proton beam size may underestimate the physical size.

\begin{table*}
\caption{Comparison of hemisphere focusing experiments. We emphasize the variability in both characterization technique and proton energy of interest for a direct quantitative comparison.}
\begin{ruledtabular}
\begin{tabular}{l | cccc | cc | ccc | c}
\multicolumn{1}{c}{} & \multicolumn{4}{c}{Laser parameters} & \multicolumn{2}{c}{Target parameters} & \multicolumn{3}{c}{Focusing behavior} &\\
Source & E (J) & $\tau$ (fs) & $D_{L}$ ($\mu$m) & I $(10^{19}$ W$/$cm$^{2}$) & $D_{h}$ ($\mu$m)\footnote{Hemisphere diameter.} & $\Psi=D_{h}/D_{L}$ & $\Delta z$ ($\mu$m) & $R_{f}$ ($\mu$m)\footnote{Proton beam focal radius.} & $\Delta z/R_{h}$\footnote{Focal location normalized to hemi radius. A value greater than unity measures a focus outside of the hemi.} & Technique\\ 
\hline
Patel \cite{patel_isochoric_2003} & 10 & 100 & 50 & 0.5 & 320 & 6.4 & 160 & 23 & 1.00 & Heating\\
Patel \cite{patel_integrated_2005} & 400 & 400 & 7 & 50 & 1000 & 143 & 500 & 47 & 1.00 & Heating\\
Snavely \cite{snavely_laser_2007} & 170 & 700 & 50 & 0.3 & 360 & 7.2 & 350 & 103 & 1.94 & Heating\\
Offerman \cite{offermann_characterization_2011} & 80 & 600 & 7 & 20 & 800 & 114  & 657 & 189 & 1.64 & Mesh rad.\\
Kar\footnotemark[4] \cite{kar_ballistic_2011} & 350 & 750 & 10 & 40 & 700 & 70  & 800 & 25 to 90 & 2.29 & Mesh rad.\\
Bartal \cite{bartal_focusing_2012} & 75 & 550 & 90 & 0.1 & 600 & 6.7  & 75 & 15 to 30 & 0.25 & Mesh rad.\\
Chen\footnotemark[4] \cite{chen_focusing_2012} & 1 & 320 & 11 & 0.3 & 800 & 73  & 370 & 15  & 0.93 & Proton rad.\\
\hline
 &  &  &  &  & 220 & 6.1 & 101 & 9.4 & 0.92 & \\
This work & 20 & 40 & 36 & 3.4 & 325 & 9.0 & 120 & 8.6 & 0.74 & Mesh rad.\\
 &  &  &  &  & 525 & 14.6 & 83 & 9.3 & 0.32 &
\end{tabular}
\end{ruledtabular}
\footnotetext[4]{These experiments used a semi-cylinder instead of a hemi target. \cite{kar_ballistic_2011} used a $250~\mu$m-thick target.}
\label{tab:FocusLitComp}
\end{table*}

Figure \ref{fig:pointing_divergence}(c) shows the beam divergence decreasing as a function of RCF layer. Qualitatively, protons emitted normal at the location where the laser hits are accelerated for longer distances and reach higher energies.

\section{Discussion and Conclusion} \label{sec:conclusion}

A central thrust of this experiment was characterizing the proton beam focal location using mesh radiography. However, this diagnostic only accesses the virtual source properties inferred by projecting straight line trajectories from the RCF image back through the mesh plane. To circumvent this limitation, we assumed that the protons travel in a hyperbolic trajectory and leveraged how the virtual focus position changes with mesh distance $L_m$ to infer a physical source (Figs. \ref{fig:dz_vs_distance} and \ref{fig:beam_evol}). However, an  alternative explanation of the changing virtual focus with mesh distance is that the mesh became charged from fast electrons and x-rays ahead of the proton beam; meshes placed closer to the interaction could become charged to a higher degree and would have less time to discharge before protons arrived, resulting in a stronger proton deflection.

However, the role of mesh charging was investigated and found to be minimal in this experiment due to the small degree of charging and fast discharge rate. Through auxiliary PIC simulations, we estimate $\sim50~$pC of uncompensated electron charge preceding the main proton beam that could charge the mesh negative. Previous experiments and theory have shown that the electrical discharge speed is close to the speed of light and would discharge the mesh exponentially \cite{ahmed_investigations_2016,poye_dynamic_2015}. The characteristic discharge time is estimated as $\tau=R_{mesh}/c=1.5~$ps where $R_{mesh} \approx 500~\mu$m is the radius of the mesh that the proton interacts with. For the worst case scenario, the closest mesh is located at $L_m = 280~\mu$m, which has an 8 ps time of flight for 5 MeV protons, associated with Layer 3 in the mesh. This mesh discharges from 50 pC to 0.2 pC in this time duration. However, to reproduce the observed trend in virtual source, we estimate the mesh must be charged to $\sim5~$pC and it is therefore unlikely that the mesh charging influenced the proton trajectories at the diagnosed energies of $\sim5~$MeV. 

Another indication that the mesh charging does not impact our measurements is the observed energy dependence of the virtual focus. If the mesh were significantly charged, it would alter the virtual focus in a way that depends on both the mesh distance and the proton energy. A mesh placed close to the interaction would acquire more charge and thus produce a stronger energy dependence on the virtual focus since the electric deflection is inversely proportional to the proton energy, $d_E\propto E_p^{-1}$. In contrast, a farther mesh would become less charged and have a weaker dependence on proton energy. Instead, we observe that the variation in virtual source with proton energy is constant for both close and far meshes, suggesting that the mesh does not impact the proton beam dynamics, as can be seen from Fig. \ref{fig:dz_vs_layer}.

In this experiment, we find that the normalized focal location $\Delta z _{phys}/R_h<1$ and decreases with increasing hemisphere size, moving the focus deeper inside the hemisphere (Fig. 5). Larger hemis (i.e. larger $\Psi$ values) behave more like flat foils and have degraded focusing, with the focus approaching the target surface as $\Delta z _{phys}/R_h\rightarrow0$. In our measurements, $\Delta z _{phys}/R_h=0.92$ for the 220$~\mu$m hemi $(\Psi=6.1)$ and decreases down to $\Delta z _{phys}/R_h=0.32$ for the 525$~\mu$m hemi $(\Psi=14.6)$. While smaller hemis were not tested due to manufacturing limits, a smaller target may produce further improved focusing with $\Delta z _{phys}/R_h>1$ under these experimental conditions. However, there is a clear tradeoff between $\Psi$ and pointing stability, as shown in Fig. \ref{fig:pointing_divergence}(a). It is unclear if the proton beam pointing is correlated with a transverse shift in the physical proton focal location, as discussed in simulations in Ref. \cite{chen_focusing_2012}.

We also find that statistics enabled by an enhanced shot rate are important to reliably detect trends in the data. With over 70 shots, we could establish systematic trends across the experimental variables, rather than relying on a few shots like prior works. This is the first experiment to measure proton focusing from hemispheres at an enhanced shot rate. The large sample size is critical to informing scaling laws to the higher driver energies required for proton fast ignition.

Table \ref{tab:FocusLitComp} compares our results with prior hemisphere and semi-cylinder experiments that diagnosed the focusing behavior with either secondary heating of a sample, mesh radiography, or side-on proton radiography. Due to the variability in laser parameters, target geometries, and diagnostic methods, direct quantitative comparison is difficult. Nevertheless, reducing the parameter space to laser intensity, dimensionless focusing geometry $\Psi=D_{hemi}/D_L$, and dimensionless focal location $\Delta z/R_h$ offers interesting points of comparison. Firstly, we infer $\Delta z/R_h$ between 0.32 and 0.92 which spans from inside the hemi to near the geometric center. Several other experiments also report focusing at or inside the geometric center \cite{patel_isochoric_2003,patel_integrated_2005,bartal_focusing_2012,chen_focusing_2012}. These experiments vary across a wide range of intensities and $\Psi$ values but are generally at lower intensities $I<10^{19}~$W/cm$^2$ and smaller $\Psi<10$. These contrast with the experiments that predict a focus downstream of the hemi center \cite{snavely_laser_2007,offermann_characterization_2011,kar_ballistic_2011} and generally have larger intensities $I>10^{20}~$W/cm$^2$ and larger $\Psi>70$. Physically, the intensity roughly controls the electron temperature while $\Psi$ shapes the electron sheath and focusing fields. The focus behavior presented here is qualitatively consistent with the simulation results from Ospina-Bohórquez \emph{et al.} \cite{ospina-bohorquez_scaling_2025} which showed optimal focusing for $\Psi = 6$ to 8.5 under similar conditions and degraded focusing for $\Psi=13.5$. However, that study predicted proton channeling and a focus outside of the hemisphere at optimal $\Psi$ value. Target thickness also plays a role, as in Kar's experiment where the 250$~\mu$m-thick target likely reduced electron temperature and $\Psi$ value since the electron beam diverges inside the solid, effectively increasing the equivalent laser spot size. The reported energy-dependent trends of downstream-shifted focal locations and slightly smaller virtual focal size are consistent with previous work \cite{bartal_focusing_2012,kar_ballistic_2011}.

In conclusion, we performed the first systematic study of proton focusing from laser-irradiated hemispherical targets of different diameters. A detailed mesh radiography analysis was performed over 70 shots, due to advances in enhanced shot rate targetry. Mesh radiography measurements at several distances allowed us to constrain both the virtual and physical focusing properties. We find that the physical focus is near the geometric center for the smallest diameter hemi ($\Psi=6.1)$, but shifts deeper inside the hemisphere for larger diameter targets ($\Psi=14.6)$. A mesh transition analysis measures the virtual focal spot size to be $9\pm3~\mu$m for all hemisphere sizes. Lastly, the proton beam pointing was characterized and found to be more erratic for small hemispheres as laser pointing stability became more impactful. These results provide critical benchmarks for scaling proton focusing toward the high driver energies demanded by proton fast ignition.

\section*{Acknowledgments}
The authors thank the Colorado State University Advanced Beam Laboratory laser facility staff for their help in conducting these experiments as well as the targetry team at Focused Energy for providing the targets and developing the enhanced rep-rate hemisphere alignment system. The research described in this paper was funded under the INFUSE program – a DOE SC FES public-private partnership – under CRADA No. 2725 between Princeton Plasma Physics Laboratory and Focused Energy Inc. company under its U.S. Department of Energy Contract No. DE-AC02-09CH11466. The experimental work was supported by the U.S. Department of Energy’s (DOE) Office of Science (SC) Fusion Energy Sciences (FES) program under DE-SC0021246: the LaserNetUS initiative at the Advanced Beam Laboratory. The PPPL team's experimental activities, numerical modeling and calculations supporting the experimental results were conducted under the Laboratory Directed Research and Development (LDRD) Program at Princeton Plasma Physics Laboratory, a national laboratory operated by Princeton University for the U.S. Department of Energy under Prime Contract No. DE-AC02-09CH11466. The United States Government retains a non-exclusive, paid-up, irrevocable, world-wide license to publish or reproduce the published form of this manuscript, or allow others to do so, for United States Government purposes.

\bibliography{PFocusing}

\end{document}